# The Size and Nature of Lyman $\alpha$ Forest Clouds Probed by QSO Pairs and Groups

Yihu Fang[1], Robert C. Duncan[2], Arlin P. S. Crotts[1] and Jill Bechtold[3]

## ABSTRACT

Closely separated QSO pairs and groups make it possible to probe the size, geometry and spatial clustering of Ly$\alpha$ forest clouds. Recent spectroscopic observations of Q1343+2640A/B give evidence that the transverse sizes of Ly$\alpha$ clouds are very large at redshifts $\sim 2$ (Bechtold et al. 1994, Paper I). In this paper, we describe a robust Bayesian statistical method for determining cloud sizes in spherical and in thin disk geometries, apply this method to the available data, and discuss implications of our results for models of Ly$\alpha$ clouds.

Under the assumption of a population of uniform-size and unclustered Ly$\alpha$ clouds, the data from Q1343+2640A/B give a 99% confidence lower and upper bounds $61 < R < 533$ $h^{-1}$ kpc on the radius of spherical clouds at $z \approx 1.8$, with a median value of 149 $h^{-1}$ kpc $[(\Omega_0, \Lambda_0) = (1, 0)$, and $h \equiv H_0/100$ km s$^{-1}$ Mpc$^{-1}]$. The baryonic mass of such large clouds, if they are roughly homogeneous and quasi-spherical, is comparable to the baryonic mass of dwarf irregular galaxies. Their cosmic overdensity is close to the turn-around density but generally below the virialization density, suggesting a population of gravitationally bound but unvirialized protogalactic objects at $z \approx 2$. The comoving volume density of these clouds is similar to that of the faint blue galaxies (FBGs) at the limiting magnitude $B \sim 26 - 27$, if these FBGs are distributed approximately over the range of redshift from 0.8 to 2. The timescale for dynamical collapse of overdensities like these clouds is also comparable with the cosmic time difference between $z \sim 2$ and $z \sim 1$. Both populations of objects show similar weak clustering in space. All this evidence suggests a possible identification of Ly$\alpha$ clouds as the collapsing progenitors of the FBGs at $z \sim 1$.

We also investigate the other closely separated QSO pairs with published high-quality spectra: Q0307-1931/0307-1932, Q0107-0232/0107-0235, and the triplet of Q1623+268. Imposing an uniform $W_0 \geq 0.4$ Å counting threshold on all the linelists, we find a trend of larger inferred cloud radius with larger proper separation of QSO pairs, significant at the 3.4 $\sigma$ level. This indicates that the idealization of unclustered, uniform-sized clouds does not accurately describe the Ly$\alpha$ cloud population. Present data are insufficient to resolve with confidence whether this effect is due to clustering or non-uniform cloud size. There is a suggestion, however, that at low redshifts a residual population of larger clouds remains.

---

[1]Dept. of Astronomy, Columbia University, 538 W. 120th Street, New York, NY 10027

[2]Dept. of Astronomy, University of Texas, Austin, TX 78712.

[3]Steward Observatory, University of Arizona, Tucson, AZ 85721

*Subject headings:* cosmology: observation - galaxies: intergalactic medium - QSOs: absorption lines

## 1. INTRODUCTION

Since Ly$\alpha$ forest absorption lines were first observed in QSO spectra (Lynds 1971), they have been discovered everywhere from the nearby universe to the redshift of the most distant QSOs known. The origin and nature of these Ly$\alpha$ forest absorbers is unclear, although many models and formation scenarios have been proposed. Progress in understanding the clouds has been slow because most observational studies only probe absorber structure in the velocity space along the line of sight, yielding almost no information on the sizes and shapes of the absorbers. From such observations it is impossible to estimate physical properties such as the density and ionization state of the absorbing gas, the absorbers' masses, and their number density in space.

Closely separated QSO pairs and gravitationally lensed QSOs provide two or more adjacent raypaths, allowing one to sample the clouds' extents in the transverse direction. Gravitationally lensed QSOs usually have raypath separations spanning significantly subgalactic scales in the observed Ly$\alpha$ forest; studies of these spectra have shown that the Ly$\alpha$ clouds are much larger than a few kiloparsecs (Weymann & Foltz 1983; Smette et al. 1992). QSO pairs with separation larger than a few arcminutes, corresponding to proper separation up to the order of megaparsecs, have the potential to study the large scale spatial distribution of Ly$\alpha$ or metal absorbers (Crotts 1985; Jakobsen et al. 1986; Tytler et al. 1993; Elowitz et al. 1995). For component separations of a few arcminutes or less, multiple raypaths may probe the physical size of the absorbers or their smallest clustering scale (Foltz et al. 1984; Shaver & Robinson 1983; Crotts 1989).

Our recent spectroscopic observations of Q1343+2640A/B, with an angular separation of 9.5″ and $z = 2.03$ revealed that many Ly$\alpha$ absorption lines are in common to both spectra, and so determined that the Ly$\alpha$ cloud size is of the order of $\sim 100\ h^{-1}$ kpc; much larger than previously thought by most workers (Bechtold et al. 1994, Paper I; Crotts et al. 1994, Paper II). An independent measurement of the same pair obtained similar results (Dinshaw et al. 1994). These studies significantly constrain Ly$\alpha$ cloud models (Paper I). In particular, some versions of pressure-confined and freely expanding Ly$\alpha$ cloud models are now excluded, as are models of stably-confined gas concentrations trapped in the gravitational potential wells of spherical cold dark matter (CDM) minihalos.

More extensive spectroscopic studies of QSO pairs with various angular separations will be undertaken in the near future. Thus robust methods of statistical analyses are needed to determine size bounds. In this paper (§2.1) we demonstrate an application of Bayes Theorem to the Ly$\alpha$ cloud size problem, as briefly outlined in Paper I, and extend this analysis to more general cloud geometries. In §2.3 we discuss the impact of cloud size constraints on Ly$\alpha$ cloud models, leading us to propose, in §3, a picture of Ly$\alpha$ clouds at $z \sim 2$ as non-linear, gravitationally contracting objects, and suggesting a connection with faint blue galaxies at $z \sim 1$. In §4, we analyze existing data in the literature of six QSO pairs with various angular separations, illustrating the limitations of the clustering-free model of clouds of a single uniform radius.

## 2. Ly$\alpha$ CLOUD SIZE MEASUREMENT FROM QUASAR PAIR SPECTRA

### 2.1. Bayesian Statistics for Cloud Size Bounds Applied to Q1343+2640A/B

To constrain cloud sizes, consider a binomial random process producing a number of "hits" $\mathcal{N}_h$ and "misses" $\mathcal{N}_m$ in two adjacent lines of sight, with proper separation of raypaths of $S$. A "hit" occurs when the absorption line appears in *both* of the two component QSO spectra, while a "miss" occurs when the line appears only in the spectrum of component A *or* in the spectrum of component B. In practice, we must impose some selection criteria to the linelists, as described in Paper I. By imposing a signal-to-noise ratio cutoff of 3.5 $\sigma$ on the linelist and $\Delta v < 150$ km s$^{-1}$ between the two common lines, we get $\mathcal{N}_h = 11$ and $\mathcal{N}_m = 4$. Given the observation of $\mathcal{N}_h$ hits and $\mathcal{N}_m$ misses, with the probability for a hit given by $\psi$, the *likelihood function* of this binomial process is $\mathcal{L}(\mathcal{N}_h, \mathcal{N}_m | \psi) = \psi^{\mathcal{N}_h}(1-\psi)^{\mathcal{N}_m}$.

If we idealize the Ly$\alpha$ forest absorbers as uniform-radius spherical clouds, we can also calculate the probability $\phi$ that a second raypath intersects a cloud given that the first raypath already does. With the definition $X = S/2R$, where $S$ is the proper separation of raypaths and $R$ is the cloud radius, we find (McGill 1990)

$$\phi = (2/\pi) \left[\cos^{-1} X - X(1-X^2)^{1/2}\right] \qquad \text{for} \quad X < 1, \tag{2.1}$$

and $\phi = 0$ otherwise.

There is some misunderstanding in the literature that the two probabilities $\psi$ and $\phi$ are the same in a uniform-spheres cloud model (McGill 1990; Smette et al. 1992; Paper I; Dinshaw et al. 1995). For a random placement of clouds along the two lines of sight, $\psi$ is the probability that both raypaths intersect the cloud given that *at least* one raypath does. Thus $\mathcal{N}_m$, the number of negative outcomes to the binomial process with probability $\psi$, is defined as the total number of unpaired lines in *both* spectrum A and B. The function $\phi$, on the other hand, is the probability that one line of sight intersects a spherical cloud, given that the other adjacent raypath already does. This function would be the appropriate theoretical probability for the binomial random process with observed realization $\{\mathcal{N}_h, \mathcal{N}_m\}$ only if we redefined $\mathcal{N}_m$ as the number of unpaired lines in spectrum A but not B (or alternatively, B but not A).

The two probabilities are simply related by

$$\psi = \frac{\phi}{2 - \phi}, \quad (2.2)$$

if the two raypaths have approximately the same propensities for detectable intersections with Ly$\alpha$ clouds. In practice, line detection thresholds can vary from one spectrum to the other and/or across a single spectrum because of the intrinsic magnitude difference of QSOs, their intensity variations with wavelength, and the spectrograph response function (see Fig. 1c of Paper I). Thus, given a line in spectrum A, the chance of detecting a corresponding line in spectrum B, $\phi_A$, is different from the chance $\phi_B$ for detecting a line in spectrum A, given a line in B. That is, the probability $\phi$ is line-of-sight dependent because of different intrinsic signal-to-noise (S/N) ratio in the two component spectra. Moreover, both $\phi_A$ and $\phi_B$ actually vary with wavelength, as a function of the local spectral S/N values. To calculate $\phi_A$ and $\phi_B$ as a function of the S/N using our linelist detection criteria is a non-trivial problem which is sensitive to details of the cloud structure. It also depends on the equivalent widths $W_A$ (or $W_B$) of the first-drawn lines. Finally, at this level of realism one should take into account the fact that the cloud population is actually non-uniform (§4). Our current theoretical understanding of the Ly$\alpha$ clouds is too tentative, and the observational constraints too sparse, to justify such detailed and elaborate theoretical fits of the data. Thus, we will simply use $\psi$ of eq. (2.2) to describe the binomial process throughout this paper, which is tantamount to idealizing the component spectra as uniform. This gives a reasonable estimate of the (presumed uniform) cloud sizes without invoking many uncertain theoretical constructs. In §4, we further discuss the size distribution of Ly$\alpha$ clouds.

To calculate statistical bounds on cloud sizes, we use Bayes' theorem (e.g., Press 1990), which yields the *a posteori* probability density in $R$. We adopt a uniform prior distribution $f(\psi)$ that all values of $R$ are equally likely, since previous observations of lensed QSOs and QSO groups probed only much smaller and much larger spatial scales, giving only extreme lower and upper bounds to the cloud size. If these bounds are included in the prior distribution, they do not significantly affect the results.

Bayes' theorem says that the *a posteori* probability density is simply $\mathcal{P}(\psi|\mathcal{N}_h,\mathcal{N}_m) = \mathcal{L}(\mathcal{N}_h,\mathcal{N}_m|\psi)\,f(\psi)$. Thus, after being converted to the variable $R$, the probability density as a function of $R$ is

$$\mathcal{P}(R) = \frac{\psi^{\mathcal{N}_h}(1-\psi)^{\mathcal{N}_m}(d\psi/dR)}{\int_0^\infty \psi^{\mathcal{N}_h}(1-\psi)^{\mathcal{N}_m}(d\psi/dR)dR}$$

(e.g., §2.5 in Press 1990). The normalization integral in the denominator can be evaluated, yielding

$$\mathcal{P}(R) = \frac{(\mathcal{N}_h + \mathcal{N}_m + 1)!}{\mathcal{N}_h!\mathcal{N}_m!} \left(\frac{d\psi}{dR}\right) \psi^{\mathcal{N}_h} (1-\psi)^{\mathcal{N}_m}. \quad (2.3)$$

This equation and eq. (2.2) apply to a uniform cloud population, but are otherwise model–inspecific. Specializing to the case of spherical clouds [eq. (2.1)], the relevant derivative in eq. (2.3) is

$$\frac{d\psi}{dR} = \left(\frac{4}{\pi}\right) \frac{X}{R} (1-X^2)^{1/2} \frac{2}{(1-\phi)^2}.$$

It is clear from eq. (2.3) that the probability density $\mathcal{P}(R)$, and thus the statistical bounds on cloud sizes, are sensitive to the values of both $\mathcal{N}_h$ and $\mathcal{N}_m$; i.e., the results depend not only on the fraction of lines which are "hits," $\mathcal{N}_h/(\mathcal{N}_h + \mathcal{N}_m)$, but also on the total number of lines that are observed. This has not been recognized in all recent Ly$\alpha$ cloud size studies.

The probability density $\mathcal{P}(R)$ in eq.(1) of Paper I should be replaced by the correct form of eq. (2.3), with eq. (2.1) and eq. (2.2). (In Paper I we made the mistake of assuming $\psi = \phi$). In Table 1, we recalculate the lower and upper bounds of the cloud size with 99% statistical significance by choosing various signal-to-noise ratio cutoffs and including or excluding a possible BAL region as in Table 1 of Paper I. The results are still insensitive to the different line-counting criteria (Paper I), however, the median sizes increase about 70% from our previous estimates, while the upper size bounds increase by almost a factor of two. The $3.5\sigma$ cutoff sample gives a 99% confidence lower and upper bounds $61 < R < 533\ h^{-1}$ kpc, with a median value of $149\ h^{-1}$ kpc and a single most probable value (i.e., mode of the $R$-distribution) of $116\ h^{-1}$ kpc in a $(\Omega_0, \Lambda_0) = (1, 0)$ universe. For other cosmological models, these numbers scale with $S$ [e.g. larger by 1.44 for (0.1,0); larger by 1.85 for (0.1, 0.9) at $z = 2$].

As an alternative to spherical Ly$\alpha$ clouds, we now consider another idealized geometry: thin slabs (Charlton, Salpeter & Hogan 1993). Following McGill (1990), we will idealize these as circular disks, with uniform radius $R$ which is much larger than their thickness, and observed inclination angle $\theta$. The probability that one raypath intersects the disk within the angle $\theta \to \theta + d\theta$, given the other raypath already does, is

$$\phi(\theta) = \left(\frac{\cos\theta}{\pi}\right)\left[\cos^{-1}\left(\frac{X}{\cos\theta}\right) - \left(\frac{X}{\cos\theta}\right)\left(1 - \left(\frac{X}{\cos\theta}\right)^2\right)^{1/2}\right] \qquad \text{for}\ \ X < \cos\theta, \qquad (2.4)$$

and $\phi(\theta) = 0$ otherwise. By integrating over $\theta$ for the randomly oriented disks, the probability is then given by (McGill 1990),

$$\phi = \int_{-\pi/2}^{\pi/2} \phi(\theta)\ d\theta$$

Using eqs. (2.1), (2.3) and (2.4), we can have calculated $\mathcal{P}(R)$ for the circular disk model. The 99% lower and upper bounds on cloud size in the disk model are 87 and $759\ h^{-1}$ kpc respectively, with a median value of $224\ h^{-1}$ kpc and a mode of $171\ h^{-1}$ kpc in a $(\Omega_0, \Lambda_0) = (1, 0)$ universe.

We compare the probability density $\mathcal{P}(R)$ for these two models in Fig. 1a, and the statistical bounds on cloud sizes in Fig. 1b. In §4.4 we consider the case of extremely prolate spheroids, or "filaments" (e.g. Cen et al. 1994; Petitjean, Mücket & Kates 1995).

### 2.2. Equivalent Width and Velocity Differences between Common Lines

We now discuss the information about Ly$\alpha$ cloud structure that can be gleaned from comparisons between absorption lines that are observed at nearly the same redshifts in two component spectra. This corresponds to probing different regions (separated by transverse distance $S$) in the same Ly$\alpha$ cloud.

Table 1.
Line Coincidences and Size Bounds for Various
Ly $\alpha$ Forest Samples from Q1343+2640A/B

| Sample: $S/N$ Threshold and $\lambda$ Range | No. of Hits | No. of Misses | Cloud Radius Limits ($h^{-1}$kpc) (99% Confidence) | | |
|---|---|---|---|---|---|
| | | | Lower | Upper | Median |
| 3.5 $\sigma$ | 11 | 4 | 61.5 | 533.5 | 149.0 |
| 3.5 $\sigma$ (no BAL) | 6 | 1 | 58.5 | 914.0 | 213.5 |
| 4.0 $\sigma$ | 9 | 4 | 53.0 | 460.0 | 127.5 |
| 4.0 $\sigma$ (no BAL) | 4 | 1 | 44.0 | 862.5 | 159.5 |
| 4.5 $\sigma$ | 9 | 3 | 58.5 | 644.0 | 155.5 |
| 4.5 $\sigma$ (no BAL) | 4 | 1 | 44.0 | 862.5 | 159.5 |
| 5.0 $\sigma$ | 7 | 3 | 49.0 | 548.5 | 128.0 |
| 5.0 $\sigma$ (no BAL) | 3 | 1 | 37.0 | 816.5 | 131.5 |

Table 2.
K-S Probability of the deviation of $D_W$ and $D_v$
from Gaussian distribution for Q1343+2640A/B

| Sample: | No. of Lines | K-S Probability | |
|---|---|---|---|
| | | $D_W$ | $D_v$ |
| Paper I, all lines | 11 | $8 \times 10^{-4}$ | $2 \times 10^{-3}$ |
| Paper I, pure Ly$\alpha$ | 5 | 0.28 | 0.02 |
| Dinshaw et al. 1994 | 10 | $4 \times 10^{-4}$ | $2 \times 10^{-5}$ |

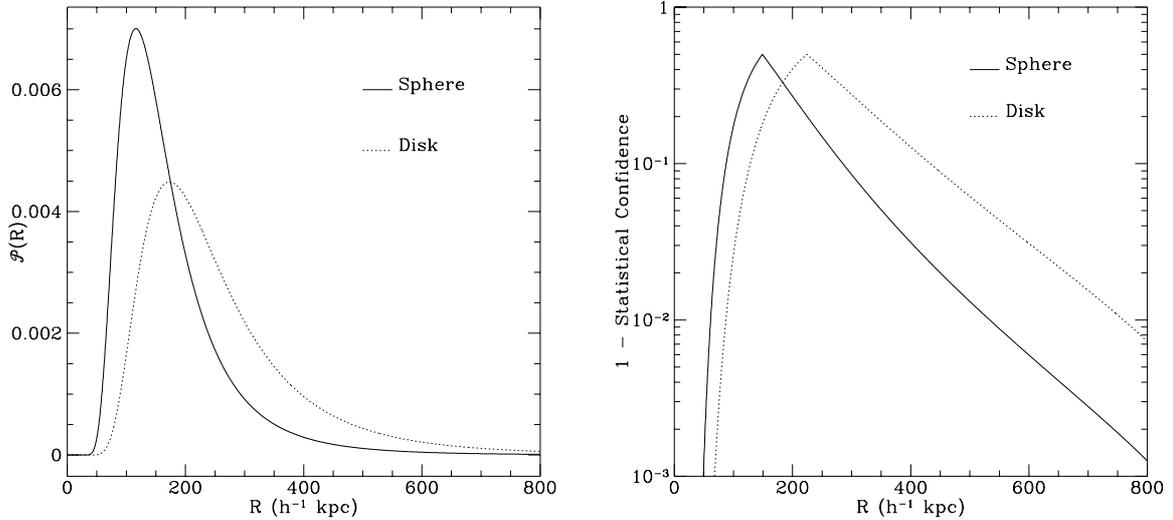

Fig. 1.— (a) Probability density for Ly$\alpha$ cloud radius, $\mathcal{P}(R)$, inferred from absorption lines in the spectra of Q1343+264A/B with the idealization of uniform spherical clouds (solid line) or uniform, thin, circular disk clouds (dash line) in an $(\Omega_o, \Lambda_o) = (1,0)$ model universe. This figure supercedes Fig. 2 of Paper I because of revisions to the Bayesian statistical analysis, as described in the text. (b) Upper and lower statistical bounds on Ly$\alpha$ cloud sizes at $z \approx 2$, obtained by integrating the probability density $\mathcal{P}(R)$ in Fig 1a. This is for an $(\Omega_o, \Lambda_o) = (1,0)$ universe; in other cosmological models the bounds simply scale with $S$, the proper separation of raypaths in the Ly$\alpha$ forest.

In Fig. 2a and 2b, we compare the rest frame equivalent widths $W_A$ and $W_B$ along the two lines of sight of Q1343+2640A/B (from Table 1 of Paper II). There are statistically significant differences in the equivalent width $W_A$ and $W_B$, in marked contrast to case of gravitational lensed QSOs with $S \leq 1h^{-1}$ kpc (e.g. Smette et al. 1992). It is clear from Fig. 2b that there exists a trend that high $W$ lines tend to have larger equivalent width differences $\Delta W \equiv |W_A - W_B|$, with a linear correlation coefficient between Max($W_A, W_B$) and $\Delta W$ of 0.92. Non-zero $\Delta W$ values could evince intrinsic transverse density gradients in the Ly$\alpha$ clouds. However, it is also possible that blending with metal lines has contaminated some of our measurements of HI lines. In the 11 pairs of common lines, six of them have partial contributions from possible metal lines in at least one of the pair of equivalent width measurements (Table 1 of Paper II). For the five pure Ly$\alpha$ lines the $W_A$ and $W_B$ correlate better than the other six (Fig. 2a). This could indicate that the clouds with the most nearly-primordial composition are homogeneous on the scale of 40 $h^{-1}$ kpc.

We have also investigated the distribution of velocity differences between the lines by cross-correlation of the two spectra. Ten pairs of lines are within $\Delta v < 150$ kms$^{-1}$, while the number of pairs which randomly locate within a bin of 150 km s$^{-1}$ is expected to be $\approx 0.9$. The r.m.s. velocity difference of these common lines are $\approx 76$ km s$^{-1}$ and $\approx 70$ km s$^{-1}$ for the five pure Ly$\alpha$ lines with typical uncertainty of measurement $\sim 30$ km s$^{-1}$. There is little evidence for correlations between $\Delta W$ and $\Delta v$, or between $\Delta v$ and max $(W_A, W_B)$.

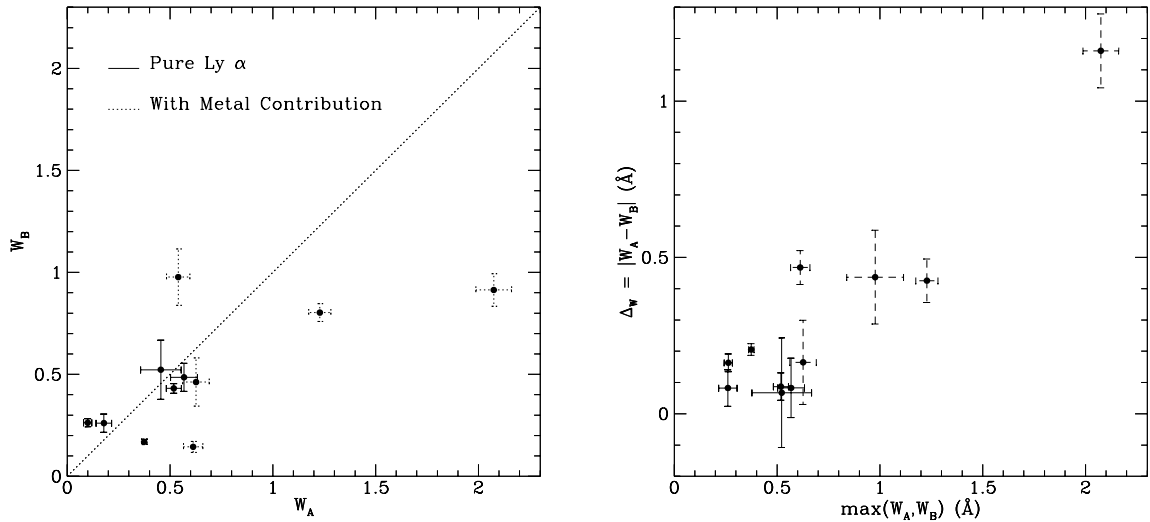

Fig. 2.— (a) In this scatter plot of data from Paper II, rest frame equivalent widths $W_A$ of Ly$\alpha$ lines in the spectrum of Q1343+266 A are plotted versus those measured in spectrum B for all common lines ("hits"). (b) The absolute value of the difference of equivalent widths between common lines in the two spectra, $|W_A - W_B|$, is plotted versus the maximum value of $W_A$ or $W_B$.

To measure how statistically significant the velocity splitting $\Delta v$ may be, we study the distribution of
$$D_v = \frac{\Delta v}{\sigma(\Delta v)} = \frac{|\lambda_A - \lambda_B|}{(\sigma_{\lambda_A}^2 + \sigma_{\lambda_B}^2)^{1/2}}.$$
In the null hypothesis of zero intrinsic $\Delta v$, the expected distribution of $D_v$ is a simple Gaussian: $P(D_v) = \mathcal{N} / (2\pi)^{1/2} \exp(-D_v^2/2)$, where $\mathcal{N}$ is the total number of matched lines. Similarly, we investigated the significance of equivalent width differences $\Delta W$ by studying the distribution of
$$D_W = \frac{|W_A - W_B|}{(\sigma_{W_a}^2 + \sigma_{W_b}^2)^{1/2}}.$$

In Fig. 3, we plot the distribution of $D_v$ and $D_W$ for our data (Paper I, II) and for the data of Dinshaw et al. (1994). A Gaussian distribution of zero mean value and unit variance is plotted on the same graphs. Both of the data sets were acquired with the Blue Channel Spectrograph at the MMT. Dinshaw et al. (1994) used an 832 line/mm grating in second order to achieve a higher spectral resolution of 1 Å FWHM, compared to our 800 line/mm grating in first order of $\sim 2$ Å

FWHM resolution (Paper I), the trade-off being that their less efficient grating setup showed no flux at wavelength less than 3400 Å and had smaller wavelength coverage.

There are significant deviations of $D_W$ from a Gaussian distribution for both data sets, which is consistent with the demonstration in Fig. 2. Deviations of $D_v$ from the Gaussian distribution are less dramatic, but still statistically significant. This is borne out by the Kolmogorov-Smirnov (K-S) probabilities for the null hypothesis of zero intrinsic $\Delta W$ and $\Delta v$, as listed in Table 2. Note, however, that the five pure Ly$\alpha$ lines show much less deviation from the Gaussian distribution both in $D_W$ and in $D_v$, although the sample is small. Thus we should be wary that the measurements of the line centroids and equivalent widths may be skewed by coincident metal lines.

The mean velocity splitting of the 11 pairs of common lines (Paper I) is $\overline{\Delta V} = 60$ km s$^{-1}$ with $\sigma(\Delta V) = 30$ km s$^{-1}$ on the scale $S \approx 40\ h^{-1}$ kpc. In the data of Dinshaw et al. (1994), with 10 pairs of common lines, $\overline{\Delta V} = \sim 48$ km s$^{-1}$ and $\sigma(\Delta V) = 16$ km s$^{-1}$. If it is truly intrinsic to the Ly$\alpha$ clouds, it cannot be attributed to the Hubble flow which is about 20 km s$^{-1}$ on scale $S$ at $z \approx 2$ in $(\Omega_0, \Lambda_0) = (1,0)$ universe. However, it could be due to the influence of local gravitational potentials. If so, the virial theorem implies a dynamic mass concentration of $\sim 5 \times 10^9 M_\odot$ within $40\ h^{-1}$ kpc. We will discuss in detail the gravitationally collapsing model in the next section.

### 2.3. Implications for Existing Models of Ly$\alpha$ Clouds

The implications of these large Ly$\alpha$ absorber sizes for two categories of currently popular models of Ly$\alpha$ clouds are discussed in Paper I. Here we update our statistical arguments to include the revised Ly$\alpha$ cloud size bounds of §2.1, and discuss a few new issues.

One class of models involve clouds that are confined by a hot intergalactic medium (Sargent et al. 1980; Ikeuchi & Ostriker 1986) or freely expanding (Bond, Szalay & Silk 1988). In some versions of these models, clouds are postulated to form in a significantly more compressed state at $z \gg 2$, through the action of shocks or gravitational collapse before the onset of reionization of the IGM. However, if these clouds subsequently expand freely at the sound speed of $c_s = (kT/\mu)^{1/2}$, where $T \sim 10^4$ K for a photoionized gas (Black 1981), they cannot be larger than $\sim 40\ h^{-1}$ kpc by $z \approx 2$. If the expansion is impeded by an external IGM pressure, then the discrepancy of the model cloud sizes with observational bounds becomes worse.

To quantitatively test these models, we compare the cloud lower size bounds (Fig. 1b) to the free-expansion length $c_s \tau_H$, where $\tau_H$ is the Hubble time at the epoch of observation ($z \approx 2$). Adopting the conservative (in the sense of giving the models the most chance to succeed) value of temperature $T = 3 \times 10^4$ K, we calculate the free-expansion length $c_s \tau_H$ as 36 $h^{-1}$ kpc for a $(\Omega_0, \Lambda_0) = (1,0)$ cosmological model; 53.5 $h^{-1}$ kpc for $(0.1, 0)$; and 85 $h^{-1}$ kpc for $(0.1, 0.9)$. By comparing the value of $c_s \tau_H$ with the probability density distribution of cloud bounds, we find that the probability of model matching is $5 \times 10^{-6}$ for a $(1,0)$ model; $1 \times 10^{-5}$ for $(0.1, 0)$; and $5 \times 10^{-4}$ for $(0.1, 0.9)$.

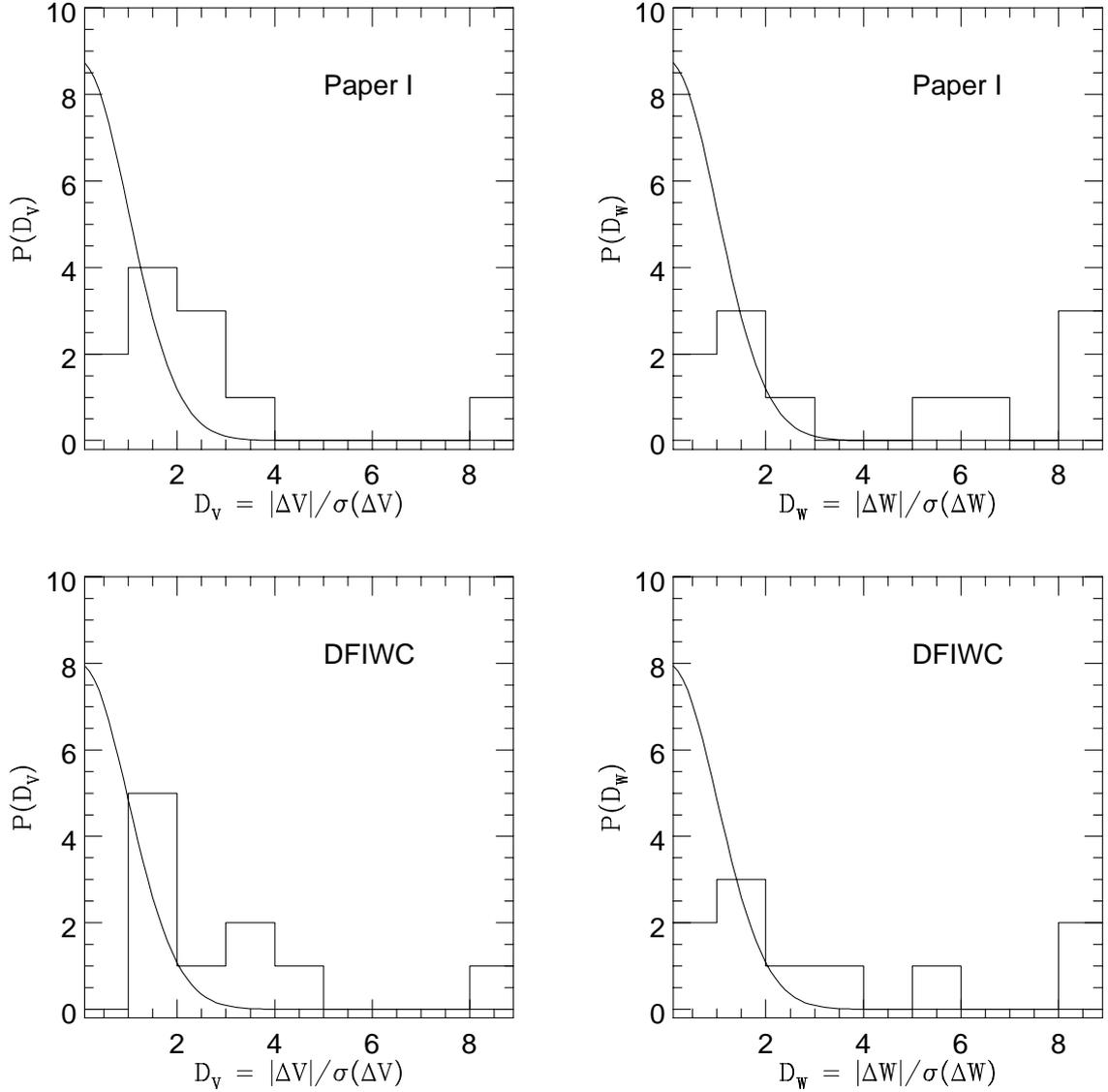

Fig. 3.— Histograms of $D_v$, the velocity differences between line-centers of common absorption lines, as measured in units of its 1-$\sigma$ measurement uncertainty, and $D_W$, the equivalent width differences, also in units of $\sigma(\Delta W)$, for our Q1343+266 A/B data (Paper II) and for the data of Dinshaw et al. (1994). A Gaussian of zero mean and unit variance is also plotted on each diagram, to illustrate how significantly observations deviate from the null hypothesis of zero intrinsic in $\Delta v$ and $\Delta W$.

Another set of models involve photoionized gas stably confined by "minihalos" of cold dark matter (Rees 1986; Miralda-Escudé & Rees 1993, "MR"). The main assumptions are (1) the matter density, dominated by CDM, has an isothermal profile, $\rho(r) = \sigma_{CDM}^2/2\pi Gr^2$; (2) the mass ratio of baryon gas to the total matter is $f_g \sim \Omega_b \sim 0.05$, as predicted by light-element nucleosynthesis in an $\Omega = 1$, $h = 0.5$ universe; (3) the gas is in ionization equilibrium with a metagalactic photon flux and finally (4) that the velocity dispersion of the hydrogen is comparable to the velocity dispersion of the CDM: $\sigma_{CDM} \approx c_s$. This last condition holds simply because if the gas were cooler than the CDM, it would be compressed into the center of the minihalo and make stars, while if it were much hotter it would escape (Rees 1986). Taken together, these four conditions predict that the impact parameter (radius $R$) at which a raypath through the cloud intercepts an H I column density $N_{14} \times 10^{14}$ cm$^{-2}$ is (MR):

$$R\,[\text{minihalo}] = 20\,\text{kpc}\,\left(\frac{N_{14}}{0.8}\right)^{-1/3}\left(\frac{T}{3\times 10^4\,\text{K}}\right)^{5/2}\left(\frac{f_g}{0.05}\right)^{2/3}\left(\frac{J_{21}}{0.3}\right)^{-1/3}. \qquad (2.5)$$

We have again adopted conservative parameter values; in particular $N_{14} = 0.8$ corresponds to the *smallest* rest-frame equivalent width line in our sample ($W = 0.22$ Å) taken with the maximum plausible thermal velocity width ($b = 30$ km s$^{-1}$). Nevertheless, the value of $R$ in eq. (2.5) is *about a factor of six times smaller than our 99% confidence lower bound, $R > 123\,(h/0.5)^{-1}$ kpc*, a circumstance which argues against the standard CDM minihalo model.

We also considered variants of the minihalo model, in two alternative cosmologies: CDM with a large cosmological constant, the "CDM-$\Lambda$" model (e.g., Efsthatiou et al. 1990); and the open-universe "Primeval Baryon Isocurvature" (PBI) model in which the collisionless halo matter is some form of condensed baryons (e.g., Cen, Ostriker & Peebles 1993). We adopted $(\Omega_0, \Lambda_0) = (0.1, 0.9)$ for CDM-$\Lambda$ and $(0.1, 0)$ for PBI. In both cases, $S$ is larger than it is for $(1, 0)$, with the size bounds on Ly$\alpha$ clouds scaling up proportionally, which *worsens* the minihalo fits. However, the mass ratio $f_g$ of gas to to the total matter can also be larger, perhaps by a large factor. The largest possible $f_g$ which can still stabilize the cloud is $f_g \sim 0.5$. (Recall that without a substantial collisionless component, a photoionized cloud is unstable [Black 1981].) Adopting this value, as well as the probable value of $N_{14} = 0.8$, $T = 3 \times 10^4$ K and $J_{21} = 0.3$, we find that a model-matching probability of cloud bound less that the value of $R$ in eq.(2.5) of 93 kpc is $1.4 \times 10^{-3}$ for CDM–$\Lambda$ model and $1.6 \times 10^{-2}$ for PBI, even if we choose a conservative Hubble constant $h = 1.0$. Thus we found no plausible scenario for dynamically-stable minihalos that was not excluded with confidence $\sim 98\%$ or stronger.

## 3. Ly$\alpha$ Clouds as Dynamically Collapsing Objects

### 3.1. The Model

We begin by considering an idealized homogeneous, ellipsoidal cloud of transverse radius or "area radius"

$$R = 100\,R_{100}\,h^{-1}\,\text{kpc}, \qquad (3.1)$$

defined so that the effective cross section is $\pi R^2$; and with a geometrical aspect ratio $\zeta$, defined so that a typical raypath through the cloud is $\zeta R$.

The neutral hydrogen density of the cloud gas is then

$$n_{\rm HI} = N/\zeta R = 3 \times 10^{-10}\, h^{-1}\, N_{14}\, \zeta^{-1}\, R_{100}^{-1}\,{\rm cm}^{-3}, \qquad (3.2)$$

for a column density of $N = 10^{14}\, N_{14}\,{\rm cm}^{-2}$. We assume that the cloud is photoionized by the metagalactic ionizing background flux, with Lyman-limit intensity given by $J_\nu = 10^{-21} J_{21}\,{\rm erg\,cm}^{-2}\,{\rm s}^{-1}\,{\rm Hz}^{-1}\,{\rm sr}^{-1}$. Ionization equilibrium implies $n_{HI}/n_{HII} = 6.6 \times 10^{-6}\, T_4^{-3/8}\, J_{21}^{-1/2}\, N_{14}^{1/2}\, \zeta^{-1/2}\, R_{100}^{-1/2} h^{1/2}$, where the temperature is $T = 10^4\, T_4$ K. The intensity of the IGM ionizing flux can be measured from the proximity effect in the Ly$\alpha$ forest (Bajtlik, Duncan & Ostriker 1988; Lu, Wolfe & Turnshek 1991; Bechtold 1994), yielding $J_{21} \sim 1$ for $1.8 < z < 4$. In particular, $J_{21}$ could be as low as 0.3, which is comparable to the integrated radiation from the observed QSO population at $z \approx 2$, or as high as $J_{21} = 3$ (Bechtold 1994). The total baryon mass of this cloud is

$$M_b = 1.3 \times 10^9\, M_\odot\, \zeta^{1/2}\, R_{100}^{5/2}\, h^{-5/2}\, N_{14}^{1/2}\, J_{21}^{1/2}\, T_4^{3/8}, \qquad (3.3)$$

comparable to the baryon mass of a dwarf irregular galaxy (Tully et al. 1978) in the interesting range $10^{-2} \leq \zeta \leq 1$.

For an $\Omega = 1$ cosmological model, in which we assume that the baryon to dark matter ratio in the cloud is comparable to the general ratio of cosmic baryon density $\Omega_b = 0.01\,(\Omega_{b,-2}) h^{-2}$ as implied by light-element nucleosynthesis, the fractional overdensity compared to the general cosmic density $\delta \equiv \Delta\rho/\rho_{cr}$ is given by

$$\delta = 8\, \Omega_{b,-2}^{-1}\, \zeta^{-1/2}\, R_{100}^{-1/2}\, N_{14}^{1/2}\, (J_{21}/0.3)^{1/2}\, T_4^{3/8}\, \left(\frac{1+z}{3}\right)^{-3}. \qquad (3.4)$$

This value is very close to the turn-around density $\delta_{turn} \approx 6$ for an isolated spherical perturbation ($\zeta \sim 1$) in an $\Omega = 1$ universe (Gunn & Gott 1972; Peebles 1980, §19), and well below the virialization density $\delta_{vir} \approx 2^3\, \delta_{turn} = 50$, a circumstance which suggests that the Ly$\alpha$ forest absorbers may be unvirialized, collapsing protogalactic clouds. This implies that the Ly$\alpha$ clouds with column densities from $10^{14}$ cm$^{-2}$ up to $N_H = 4 \times 10^{16}\, R_{100}$ would be gravitationally bound but unvirialized at the epoch of observation, $z \approx 2$. The velocity difference on the scale of 40 $h^{-1}$ kpc of $< 100$ km s$^{-1}$ (§2.2) is also consistent with the picture of gravitationally contracting clouds.

Such a collapsing cloud will ultimately virialize at a velocity dispersion,

$$v_{vir} = 100\,{\rm km\,s}^{-1}\, \Omega_{b,-2}^{1/2}\, \zeta^{1/4}\, R_{100}^{3/4}\, N_{14}^{1/4}\, J_{21}^{1/4}\, T_4^{3/16}, \qquad (3.5)$$

where $R_{100}$ still quantifies the cloud size at the turn-around epoch (e.g. $z = 2$), not the virialized object at lower redshift. Since this velocity is much larger than the sound speed in photoionized gas, $c_s = 15\ T_4^{1/2}$ km s$^{-1}$, gas pressure cannot avert continued collapse of the baryons, occurring on a dynamical time,

$$t_{dyn} = 2 \times 10^9\ \text{yr}\ h^{-1}\ \Omega_{b,-2}^{1/2}\ \zeta^{-1/4}\ R_{100}^{1/4}\ N_{14}^{-1/4}\ J_{21}^{-1/4}\ T_4^{-3/16}. \tag{3.6}$$

This gives rise to dwarf galaxy formation at redshift

$$1 + z_f = \left[\frac{3}{2}\ H_0\ t_{dyn} + (1+z)^{-3/2}\right]^{-2/3}, \tag{3.7}$$

or $z_f \approx 1$, insensitive to the redshift $z$ of the Ly$\alpha$ absorber as long as $z \geq 2$.

The measurement of the transverse size bound of the Ly$\alpha$ forest absorbers in Q1343+2640A/B allows us to estimate their comoving density without additional model assumptions. It is given by,

$$n_{L\alpha} = \frac{d\mathcal{N}/dz}{(\pi R^2)\ (d\ell/dz)\ (1+z)^3}. \tag{3.8}$$

The redshift density of Ly$\alpha$ lines with rest-frame equivalent widths $W_0 > 0.3$ Å (Lu, Wolfe & Turnshek 1991; Bechtold 1994), comparable to the limiting value of the $W$ to which our size bounds apply, is

$$\frac{d\mathcal{N}}{dz} \approx 40 \qquad \text{at} \qquad z \approx 2. \tag{3.9}$$

The proper distance $\ell$ along the line of sight, per unit redshift, is

$$\frac{d\ell}{dz} = \frac{c}{H_0}\ (1+z)^{-2}\ (1 + 2\ q_0\ z)^{-1/2}.$$

Thus,

$$n_{L\alpha} = 0.24\ h^3\ R_{100}^{-2}\ \text{Mpc}^{-3}, \tag{3.10}$$

at $z \approx 2$, for $W > 0.4$ Å, and $q_0 = 0.5$. The absorber density $n_{L\alpha}$ is smaller by only a factor of $\sim 1.5$ if $q_0 = 0.1$.

### 3.2. Structure of Ly$\alpha$ clouds

The picture of quasi-spherical and homogeneous clouds is obviously very idealized. Rather than spherical, many Ly$\alpha$ absorbers at $z \sim 2$ might be sheet-like or filament-like while still unvirialized, or consist of merging substructures which could themselves contain partially pressure-supported gas (e.g., Cen et al. 1994). Nevertheless, given their large transverse sizes (on the order of $R \sim 100\ h^{-1}$ kpc), there are several reasons for favoring the generalized picture of most Ly$\alpha$ clouds being dynamically collapsing objects, at least in a first approximation:

(1) Most overdensities drawn from a Gaussian random field first separate from the Hubble flow as quasi-homogeneous ellipsoids, which contract under the influence of the ellipsoid self-gravity and the external quadruple shear (Eisenstein & Loeb 1995, "EL"). Both forces are linear in the coordinates and therefore maintain homogeneity as the ellipsoid turns around and contracts. For a standard CDM spectrum of perturbations, most objects evolve from quasi-spherical initial states to sheet or filament geometry, and then to complete virialization, with the shape evolution determined primarily by the external shear and not the initial triaxiality (EL). Such collapsing objects pass through an extended early phase and the evolution accelerates with collapse, in which the analysis in the above section for a homogeneous and quasi-spherical ($\zeta \sim 1$) cloud would be a reasonable approximation. During this phase, the objects subtend a much larger spatial cross-section than they do at later times. This suggests the steep redshift evolution of Ly$\alpha$ forest line numbers for $1.8 < z < 3.5$ is driven by dynamical cloud contraction. This suggests that the steep redshift evolution of Ly-alpha forest line numbers for $1.8 < z < 3.5$ is driven by dynamical cloud contraction. An interesting study of this possibility in the linear approximation has been done by Bi, Ge & Fang (1995).

(2) The two point autocorrelation of the Ly$\alpha$ forest absorption lines along the line of sight show little or weak clustering on velocity scales of less than 300 km s$^{-1}$ (Sargent, et al. 1980; Webb 1987; Ostriker, Duncan & Bajtlik 1988; Webb & Barcons 1991; Rauch et al. 1993). On larger velocity scales, the spatial distribution of Ly$\alpha$ absorbers at $z \approx 2$ does not have the same fraction of empty space such as $\sim 10$ Mpc voids in the low redshift galaxy distribution (Carswell & Rees 1987; Crotts 1987; Duncan et al. 1989). These differences between the observed distributions of Ly$\alpha$ clouds and normal galaxies suggest that most Ly$\alpha$ clouds exist in relatively isolated environments, so are not near the highest density perturbations, and thus may not have participated in much hierarchical merging at the time of observations $z \geq 2$. This is consistent with the evidence that the Ly$\alpha$ forest clouds have been metal enriched at high redshift, but are still rather metal-poor compared with normal galaxies (cf. Cowie et al. (1995) and Tytler et al. (1995)).

### 3.3. Are Ly$\alpha$ Clouds the Progenitors of the Faint Blue Galaxies?

Deep CCD surveys (Tyson 1988; Cowie et al. 1988) have found high surface densities $\sim 3 \times 10^5$ galaxies per square degree down to limiting magnitude $B \sim 27$ (e.g., Lilly, Cowie & Gardner 1991; "LCG"). These faint objects have distinctly blue spectra, which may indicate (and are certainly consistent with) active star formation at large redshifts. Because their observed shapes are distorted by gravitational lensing when their angular positions lie near galactic clusters at $z \sim 0.5$, most of these faint blue objects (FBOs) must be background objects, with redshifts $z \geq 0.9$ (Tyson, Valdes & Wenk 1990). The redshift upper limit of most of the FBOs with $B_J < 27.5$ are estimated to be less than three from the discontinuity of the galaxy continuum at the Lyman limit break (Guhathakurta, Tyson & Majewski 1990). The redshift of faint galaxies with $B_J \sim 21 - 24$ are observed to have a median value in the range of $0.3 - 0.4$ (LCG; Colless et al. 1993), which has a trend to $z \approx 1$ toward $B_J = 25$ (Tyson 1994). Several interpretations of the FBOs have been proposed; for a review with references see Tyson (1994). One attractive possibility (Efstathiou et al. 1991; Babul & Rees 1992, "BR"; Cole, Treyer & Silk 1992; and references therein) is that the FBOs are dwarf galaxies which undergo starbursts at $z \sim 1$, most of which subsequently are disrupted by supernova-driven outflows or otherwise evolve to low surface brightnesses, making them difficult to detect in later epochs.

Estimates of the comoving number densities of the faintest FBOs with $B_J \sim 26 - 27$ is very uncertain because of their unknown redshift distribution. However, if they are due to a population of objects in the starburst phase lasting from $z \sim 1.5$ to $z \sim 0.8$, then given the surface densities of FBOs $\sim 3 \times 10^5$ deg$^{-2}$ and the volume of comoving space available, one finds a comoving density of

$$n_{FBO} \sim 0.3 \ h^3 \ \mathrm{Mpc}^{-3} \qquad (3.11)$$

in an $(\Omega_0, \Lambda_0) = (1, 0)$ universe (cf. LCG Fig. 21). This result is insensitive to the lower redshift cutoff; i.e., there is no change in the first significant digit of $n_{FBO}$ if the lower $z$-cutoff is $z = 0$ instead of $z = 0.8$. The number density changes only to $n_{FBO} \sim 0.2 \ h^3$ Mpc$^{-3}$ if one adopts $0.8 < z < 2$. For a $(\Omega_0, \Lambda_0) = (0.2, 0)$ universe, with the available comoving volume of these FBOs ranging from $0 < z < 2$ to $0.8 < z < 1.5$, the number density $n_{FBO}$ changes from $\sim 0.1$ to $0.2 \ h^3$ Mpc$^{-3}$.

Comparison of $n_{FBO}$ to eq. (3.10) gives the remarkable result

$$n_{L\alpha}[z = 2] \approx n_{FBO}[z \sim 1]$$

for the transverse cloud size of $R \sim 150$ kpc, if we consider the uncertainties of the FBOs redshift distribution. This co-moving density exceeds by $\sim 30$ the density of $L_*$ galaxies in the present epoch ($z = 0$), so FBOs and Ly$\alpha$ clouds are much more numerous than other classes of extragalactic objects. If all the Ly$\alpha$ clouds at $z \approx 2$, or if all the FBOs at $z \sim 1$, have baryon masses comparable to that of eq. (3.3), the total baryon mass in these objects is still a factor of $\sim 0.1$ below the total baryon mass inferred from considerations of light element nucleosynthesis. Thus Ly$\alpha$ forest objects and/or FBOs need not dominate the cosmic baryon density. The timescale for dynamical collapse of overdensities at $z = 2$ suggested by eq. (3.4) is also comparable to the cosmic time difference between $z \sim 2$ and $z \sim 1$ (eqs. (3.6) and (3.7)). The endpoint of such collapse is likely to trigger a burst of OB star formation, suggesting that *the Ly$\alpha$ forest absorbers with $N > 10^{14}$ cm$^{-2}$ at $z \geq 2$ are the progenitors of the faint blue galaxies at $z \sim 1$.*

The two-point angular correlation function $w(\theta)$ of those faint blue galaxies with $B_J \sim 25 - 26$ shows intrinsically weaker clustering properties than the $L_*$ galaxies at $z \approx 0$ (Efstathiou et al. 1991; Neuschaefer, Windhorst & Dressler 1991; Brainerd, Smail & Mould 1995). The amplitude of $w(\theta)$ implies a spatial two-point correlation function $\xi(r) = (r/r_0)^{-1.8}$ with a comoving correlation length $x_0 \approx 2 \ h^{-1}$Mpc if the median redshift of these galaxies is $z \sim 1$ (Efstathiou 1995). If the clustering pattern maintains in comoving coordinates out to $z = 2$, this correlation scale corresponds to a velocity difference of about $\Delta v = 350$ km s$^{-1}$ in a (1,0) model universe or 200 km s$^{-1}$ for a (0,0) universe.

This clustering scale is consistent with published estimates of the small scale line-of-sight velocity clustering found in the Ly$\alpha$ forest spectra: $\Delta v = 50 - 300$ km s$^{-1}$ (Webb 1987), $\Delta v = 200 - 600$ km s$^{-1}$ from a sample of 18 QSOs between redshift of 1.8 to 3.8 (Ostriker et al. 1988), and recently the result of $\xi \approx 1$ within $\Delta v = 100$ km s$^{-1}$ at $z \sim 3.5$ from an echelle spectrum of Q0055-269 (Cristiani et al. 1995).

Babul (1991) used the distribution of line-interval size along the line of sight to compare the velocity correlation seen in the Ly$\alpha$ forest with the three dimensional spatial correlation function. By assuming a self-similar correlation function (Davis & Peebles 1977) with the comoving correlation length evolving as $x_0(1+z)^{-2/3}$, he found that the Ly$\alpha$ clouds seen in the QSO spectra (Ostriker et al. 1988) with $\langle z \rangle \approx 2.5$ have a present comoving correlation length of 0.43 $h^{-1}$Mpc, which is about 10 times smaller than that of the present $L_*$ galaxies ($x_0 \approx 5.5$ $h^{-1}$Mpc). For comparison, the comoving correlation length of FBOs at $z \sim 1$ (Efstathiou 1995), extrapolated to the present by assuming a self-similar Davis-Peebles form, is 3.2 $h^{-1}$ Mpc. This may be consistent with the gravitational collapsing picture in which the more extended Ly$\alpha$ clouds contract during collapse to form two or more FBOs, or some fraction of low mass clouds seen at high redshift drop below the detection limit. Such scenarios could explain the apparently steeper correlation function evolution than the self-similar Davis-Peebles form.

If the Ly$\alpha$ forest clouds were made up of dynamically *stable* mini-halos at $z \geq 2$ (which subsequently could be destabilized by the declining ionizing UV flux, as suggested by BR) these clouds would have $R_{100} \leq 0.2$ (cf. §2.3), and the co-moving densities of $n_{L\alpha}$ and $n_{FBO}$ would be discrepant by a significant factor. For this reason, and because of the direct evidence for large absorber sizes, in contrast to BR we favor larger-mass, later-collapsing, dynamically unstable density concentrations as candidates for "typical" Ly$\alpha$ forest absorbers/ FBOs progenitors. On the other hand, spectroscopic observations of QSO pairs with angular separation of 1 arcmin or larger at similar redshift will tighten upper bounds on the cloud sizes. If Ly$\alpha$ clouds were proven to have much larger transverse sizes of $R_{100} \geq 3 - 5$, then the comoving densities of $n_{L\alpha}$ and $n_{FBO}$ would also be significantly discrepant, which could argue against the FBO connection.

## 4. Information from QSO Pairs with Wider Separation: Clustering and Non-Uniform Distributions of Cloud Sizes

### 4.1. The Sample

To describe the Ly$\alpha$ forest clouds as a population rather than in terms of the properties of a typical cloud, we can track the change of behavior in hit/miss frequencies for different pair separations $S$. This allows us to test our operational hypothesis that the clouds are of uniform size and not clustered. In addition to Q1343+2640, several QSO pairs or groups with adjacent lines of sight containing well-studied Ly$\alpha$ forest spectra have been presented in the literature. These include Q0307-1931/0307-1932 (Shaver & Robertson 1983), Q0107-0232/0107-0235 (Dinshaw et al. 1995) and Q1623+2651A/1623+2653/1623+2651B (Crotts 1989). We do not include

Q1517+2357/1517+2356 (Elowitz et al. 1995) because we have found that in comparison to our own unpublished data on the pair the published data suffer from significant wavelength calibration errors; we will treat this pair in a later paper (Fang & Crotts 1995). In Table 3 the Ly$\alpha$ forest redshift ranges, angular separations, proper separation range ($q_0 = 1/2$; for $q_0 = 0.1$ and no cosmological constant, multiply $S$ by 1.37, 1.37, 1.15, 1.42, 1.42, 1.43, respectively), hit and miss counts, inferred 95% and 99% confidence intervals, and median predicted cloud radii (assuming unclustered, uniform-sized spheres, according to eq. (2.3) ) are shown for each of these QSO pairs. We have regularized the data sets by imposing a uniform $W_0 \geq 0.4$ Å limit for all QSOs. For the data on Q1343+2640 (Paper I), this implies discarding the spectra below 3350Å due to insufficient continuum $S/N$ in Q1343+2640B. This filtered data is then similar to that obtained by Dinshaw et al. (1994) on the same pair. This is tantamount to adopting weaker size limits than those of Table 1 and Fig. 1 (which we still consider to be reliable) but it has the advantage of imposing a uniform selection criterion on all data sets.

We find that the 150 km s$^{-1}$ hit velocity interval adopted in Paper I for Q1343+2640 is reasonable for pairs at even larger separations. Fig. 4 shows the number of $W_0 \geq 0.4$ Å pairs in each of the six QSO pairs studied here, as a function of velocity separation between the two absorption lines in adjacent sightlines. For sightline separations $S$ up to $\approx$ 500 $h^{-1}$ kpc and perhaps as large as 700 $h^{-1}$ kpc, a clustering feature for $\Delta v \lesssim 150$ km s$^{-1}$ is significant. The strength of this feature, expressed as a two-point correlation function, appears to fall with increasing $S$.

When dealing with a triplet of sightlines, as in the case of Q1623+268, we have some additional information from the relative positions of absorbers in three sightlines rather than two, as well as some enhanced correlation between hits caused by, for instance, a hit between sightlines A and B and between B and C at the same redshift implying an enhanced probability of a hit between A and C. In fact the latter circumstance occurs at only two redshifts in the Q1623+268 triplet, at $z = 2.114$ and 2.138. This is insufficient to strongly affect the statistics of the hit/miss counts, and does not provide enough information about the three-point correlation for a useful analysis. In this paper, we will treat the three pairs of sightlines within the Q1623+268 triplet as independent samples.

For the Q1623+268 triplet, there are three cases of two neighboring lines in one spectrum both landing within 150 km s$^{-1}$ of another line in an adjacent line of sight. In this case we count one "hit" and no "misses" in our cloud size estimation to avoid one line being counted as belonging to two pairs (presumably two clouds). Again, in the data for the Q1623+268 triplet as for the Q1343+2640 pair, there is a significant enhancement in line pairs at the scale of 100-150 km s$^{-1}$, even though the hit fraction $\psi$ drops to 10−20%. As noticed in Crotts (1989), the Poisson probability of finding a deviation of random distribution as large as the observed 13 pairs within $\Delta v < 100$ km s$^{-1}$ is only $\sim 8 \times 10^{-4}$. Increasing $\Delta v$ to 150 km s$^{-1}$, one sees 16 pairs, with only 3.1 expected at random. (The random expectation is computed by averaging the number of pairs in bins from 1000 to 10000 km s$^{-1}$.)

Table 3.  Ly$\alpha$ Cloud Radius Estimates from QSO Pairs

| QSO Pair | $\theta$ (″) | Ly$\alpha$ $z$ Range | $S$, Proper Separation ($h^{-1}kpc$) | $N_h$ | $N_m$ | 95% Conf Interval ($h^{-1}kpc$) | 99% Conf Interval ($h^{-1}kpc$) | Median Radius ($h^{-1}kpc$) |
|---|---|---|---|---|---|---|---|---|
| 1343+2640A/B | 9.5 | 1.756[1]-2.035 | 39-40 | 7 | 1 | 77-841 | 59-962 | 237 |
| 0307-1931/1932 | 56 | 1.690-2.122 | 226-236 | 3[2] | 14[2] | 148-347[2] | 138-421[2] | 209[2] |
| 0107-0232/0235 | 86 | 0.481-0.952 | 301-364 | 4 | 6 | 286-918 | 252-981 | 501 |
| 1623+2651A/B | 127 | 2.043-2.467 | 493-522 | 5 | 21 | 316-667 | 290-777 | 435 |
| 1623+2651A/2653 | 147 | 1.971-2.467 | 571-604 | 3 | 19 | 321-684 | 300-801 | 440 |
| 1623+2653/2651B | 177 | 2.043-2.521 | 683-721 | 2 | 27 | 357-634 | 346-721 | 450 |

[1]We impose a general $W_o \geq 0.4$Å cutoff, which implies that the spectrum for Q1343+2640 (Paper I) with $\lambda < 3350$Å must be discarded, as well as some of the linelist in Paper II.

[2]We have imposed a $\Delta v < 150\ km\ s^{-1}$ hit limit. If one counts as a hit the line pair with $\Delta v = 192\ km\ s^{-1}$, this implies $N_h = 4$, $N_m = 12$, with a median size of $R = 246\ h^{-1}kpc$, and 95% and 99% intervals of 165-434 and 151-537 $h^{-1}kpc$, respectively.

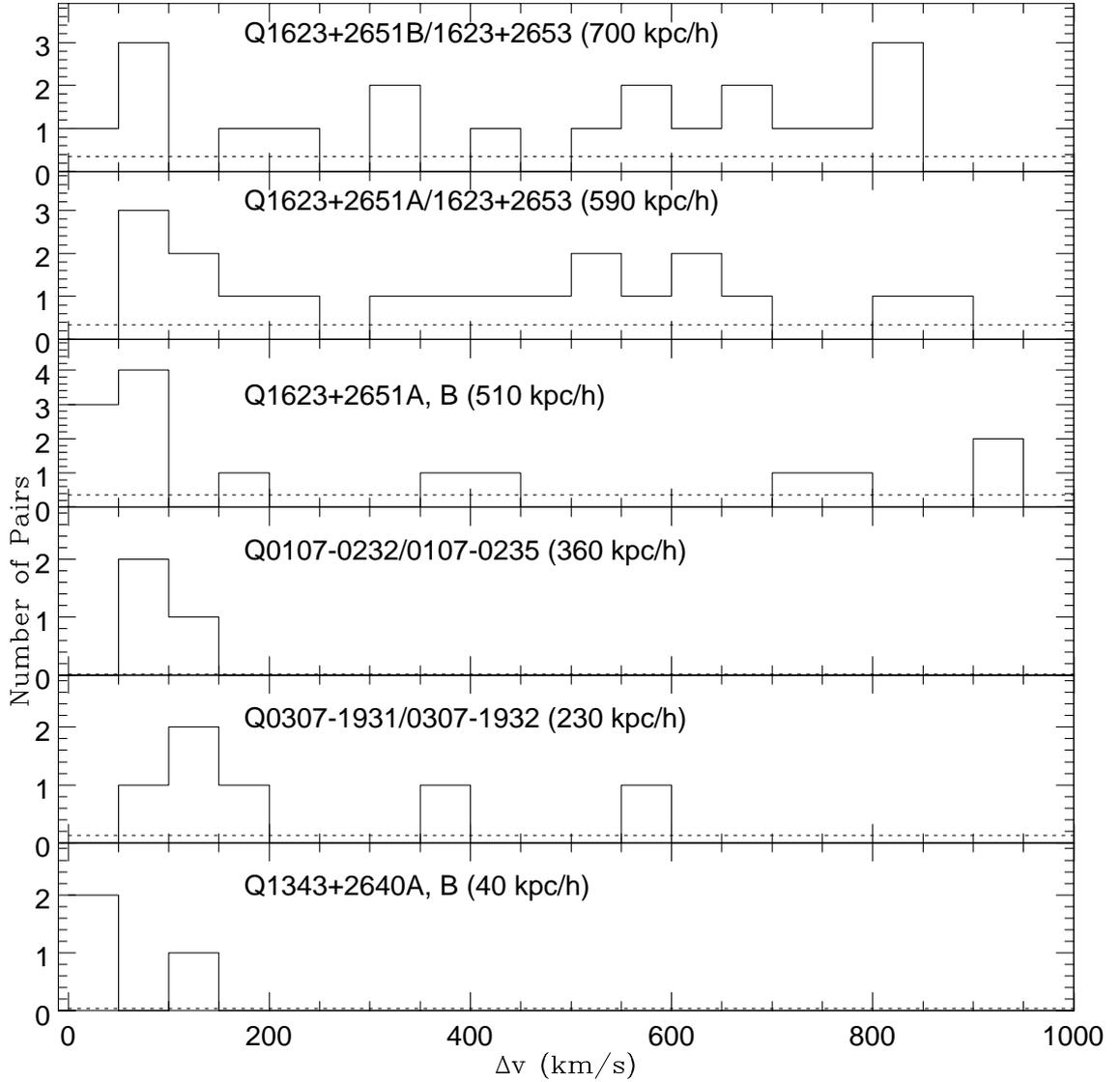

Fig. 4.— Histograms of velocity difference $\Delta v$ between pairs of Ly$\alpha$ absorption lines with $W_0 \geq 0.4$ Å in adjacent lines of sight for six QSO pairs (see Table 3). Solid lines show the observational data with a bin size of 50 km s$^{-1}$; dotted lines show the mean distribution that would be expected if the Ly$\alpha$ forests in the two lines of sight were independent and uncorrelated. The number in parenthesis for each QSO pair is the average proper separation $S$ of the adjacent lines of sight in an $(\Omega_0, \Lambda_0) = (1, 0)$ model universe.

Our original Bayesian cloud size analysis (Paper I) did not treat the accidental hits due to random velocity matches between sightlines, because this was a negligible correction for the case of Q1343+2640 A/B. Accidental hits can be dealt with by subtracting the expected number of random matches from the observed number of hits and adding twice this amount to the number of misses. However, because of the way we deal with multiple hits (as described in the preceding paragraph), a random match to a pre-existing hit does not change the number of observed hits. Hence, the number of hits are reduced by the subtraction of $N_r(1 - f_H)$, where $N_r$ is the number of random pairs expected and $f_H$ is the fraction of lines involved in hits. The number of misses is increased by $2N_r(1 - f_H)$.

### 4.2. Non-uniform Cloud Size Distribution

With these techniques in place, we can compute the cloud size $R$ for each of the QSO pairs at different separations $S$. We find an unmistakable trend of larger estimated cloud size with increasing sightline separation $S$. This is shown in Fig. 5, which plots the confidence intervals corresponding to $\pm 1\sigma$ for the various QSO pairs, along with the relation $R = S/2$, which is the lower limit of any hit detection for unclustered uniform spheres. The data for Q0307-1931/32 and the Q1623+2651A/53/51B triplet concern overlapping redshift ranges, while 95% confidence intervals in $R$ overlap only slightly or not at all. The data from the two closer pairs indicate a cloud size that is smaller and nearly inconsistent with the three larger pairs' results. This can be expressed as the slope in $R$ versus $S$ required to fit the data (the dashed lines in Fig. 5, which are nearly identical but differ depending upon our including or omitting the low-$z$ pair 0107-0232/0107-0235), which is non-zero at the $3.4\sigma$ level. (A range of constant [zero-slope] values, from 330 to 470 $h^{-1}$ kpc, peaking at 380 $h^{-1}$ kpc, are marginally consistent with all of the data, but only at a probability level of $2 \times 10^{-3}$.) Such a trend of $R$ with $S$ would not be expected if the Ly$\alpha$ clouds were truly a uniform-size, unclustered population. Note that a small but statistically significant number of hits guarantees the output of a median $R$ value slightly larger than $S/2$, with the clouds just spanning the angular gap, as is seen in Q0307-1931/32 and Q1623+2651A/53/51B where the number of hits is only 10-20% of either line sample. Thus if Ly$\alpha$ clouds are clustered, or if there exists a small sub-population of larger clouds, a trend of $R$ increasing with $S$ in our analysis would be obtained. We conclude that *Fig. 5 gives evidence for cloud clustering, non-spherical shape and/or a range of cloud sizes spanning the observed range in $S$.*

It is worth noting that the two smallest redshift pairs, 0107-0232/0107-0235 and 1343+2640A, B, fall farthest above the $R$ versus $S$ trend set by all six pairs. This can be taken as weak evidence for an increasing cloud size with cosmic time. Removing the $R$ versus $S$ dependence and expressing the remaining correlation as a power law in $1 + z$, one finds a best fit for $R \propto (1+z)^{1.89 \pm 0.98}$, or a $1.9\sigma$ result.

The trend set by the growth of $R$ with $S$ indicates that approximately 180 $h^{-1}$kpc of the discrepancy between the smaller separation pairs (1343+2640 and 0307-1931/0307-1932) and the pair 0107-0232/0107-0235 size measurements is due to the failure of the uniform size, unclustered model, and not due to evolutionary changes with redshift. There remains a suggestion, but only at the $1.6\sigma$ level, of a discrepancy of about 220 $h^{-1}$kpc between 0107-0232/0107-0235, at $z \approx 0.75$, and the trend indicated by the five $z \approx 2$ pairs, indicating that the low redshift clouds are about twice the radius of those at higher redshift. The change in $R$ with $z$ can be understood, perhaps, in the context of the large change in the number of detected clouds over the redshift range $z \approx 2$ to 0.75. The line-of-sight density $N$ of $W_0 > 0.4$ Å Ly$\alpha$ cloud detections drops by a factor of 1.7 between $z \approx 2$ (Lu, Wolfe & Turnshek 1991) and the $\langle z \rangle = 0.72$ range (Bahcall et al. 1993) of 0107-0232/0107-0235. This implies a comoving line-of-sight number density change by a factor of $\sim 1.3$ when cosmological effects are taken into account, if the cloud cross-sections are constant ($\sim 1.0$ for $q_0 = 0.1$). When one includes the increase in cloud cross-sections implied by the larger characteristic size in the 0107-0232/0107-0235 data, however, the implied comoving spatial number density has *decreased* by a factor of about 5.3 (4.0 for $q_0 = 0.1$).

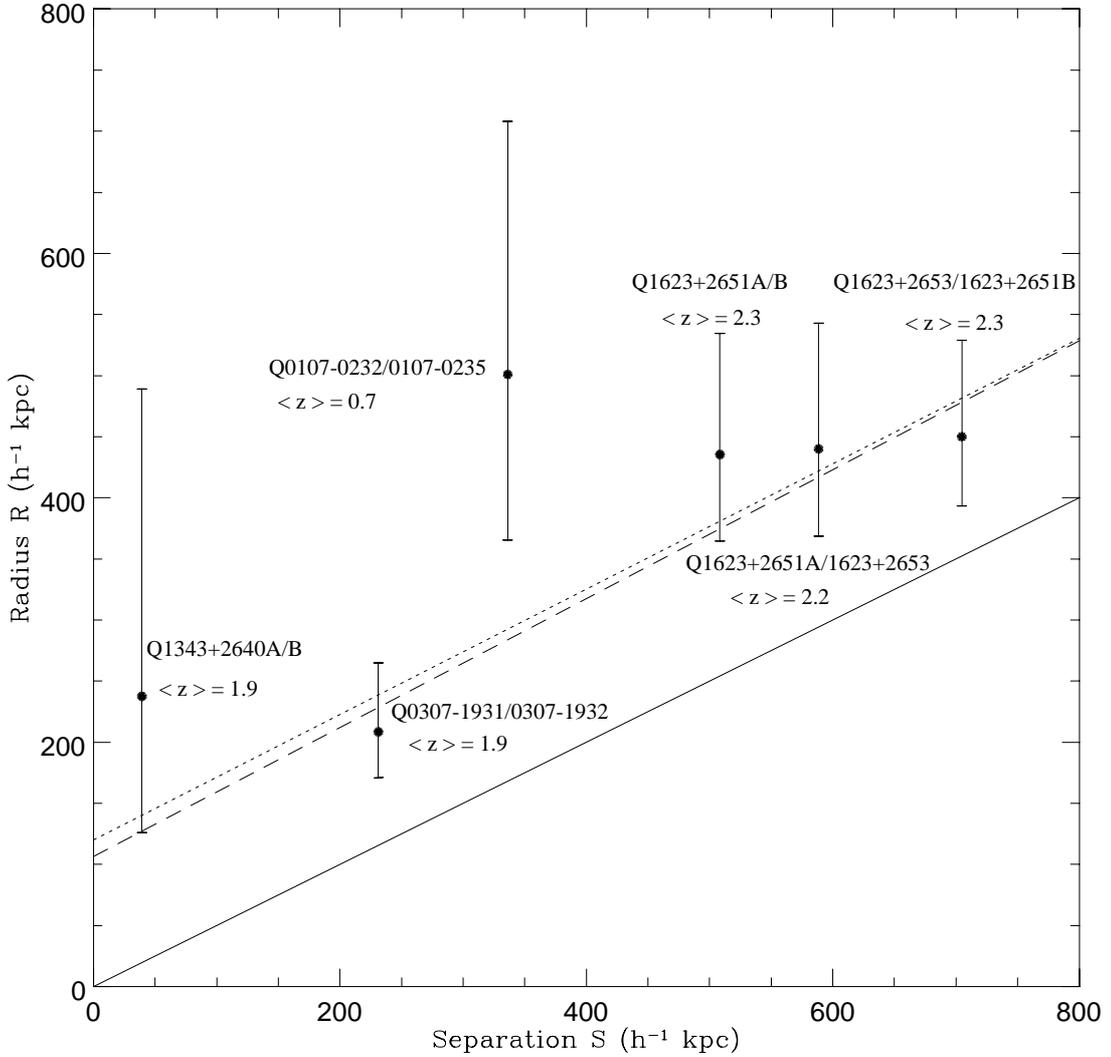

Fig. 5.— Cloud sizes $R$ estimated from the linelists of six QSO pairs, with an uniform $W_0 \geq 0.4$ Å counting threshold (Table 3), are plotted versus the average proper separation $S$ of raypaths in an $(\Omega_0, \Lambda_0) = (1, 0)$ universe, with $\pm 1\sigma$ confidence intervals. The solid line $R = S/2$ is the minimum cloud radius for any hits to occur in a population of isolated spherical clouds. The slope in $R$ versus $S$ is fitted by including (long dash line) or omitting (dotted line) the low redshift pair Q0107-0232/0107-0235.

It would appear that there is some evidence for smaller clouds disappearing from the Ly$\alpha$ sample, leaving behind a small number of larger clouds. If Ly$\alpha$ clouds, once they collapse, form structures which no longer produce Lyman-series absorption lines, this can be understood simply in terms of the fluctuation spectrum of overdensities. In the case of simple, adiabatic CDM fluctuations, for instance, the scale undergoing gravitational collapse at $z = 0.7$ is 8 times larger in proper coordinates than that at $z = 2$. From these data, we have information about the distribution of cloud sizes, and can discern whether there exist larger numbers of smaller clouds at high redshift.

Making the assumption that the clouds do not cluster, but are characterized by a power law size distribution (in spatial number density) $n(R) = kR^\alpha$ within cutoff radii $R_{min}$ and $R_{max}$, we can evaluate a simultaneous least-$\chi^2$ fit to the hit fraction $f$ for each of the five $z \approx 2$ pairs, excluding Q0107-0232/35 due to its lower redshift and marginal ($2\sigma$) deviation from the best fit to the other pairs $R$ versus $S$ dependence in Fig. 5. A binomial distribution in $N_h$ and $N_m$ is used to evaluate $\sigma_f$.

The resulting $\chi^2$ distribution in $\nu = 3$ degrees of freedom ($\alpha$, $R_{min}$ and $R_{max}$) only loosely constrains the size distribution of Ly$\alpha$ clouds. The best fit, with $\chi^2/\nu \approx 1.0$ occurs for $\alpha \approx -4.2$, $R_{min} \approx 80$ $h^{-1}$kpc and $R_{max} \approx 3$ $h^{-1}$Mpc. The $\chi^2/\nu = 2.6$ contour, corresponding to 95% confidence, extends from $\alpha \approx -6$ (and large values of $R_{min}$ and $R_{max}$, about 200 $h^{-1}$kpc and $> 3$ $h^{-1}$Mpc, respectively) to $\alpha \approx -1$ (with $R_{min} \approx 40$ $h^{-1}$kpc and $R_{max} \approx 1$ $h^{-1}$Mpc). In no case is it possible for many of the clouds to have a radius smaller than about 50 $h^{-1}$kpc. Some must be at least 400 $h^{-1}$kpc in radius (trivially, since they must span the gap between 1623+2653 and 1623+2651B). Small values of $\alpha$ (and correspondingly small values of $R_{min}$ and $R_{max}$) drive the predicted value of $\psi$ for the low-$z$ system 0107-0232/0107-0235 to within $1\sigma$ of the measured value (at the expense of driving the 0307-1931/0307-1932 $\psi$ value more than $1\sigma$ away; 1343+2640A/B is barely affected). The implications of this analysis are still unclear: it is possible to construct a size distribution where only a small fraction of the clouds, the largest ones, survive to low redshift, or alternatively where the single low-$z$ size estimate does not deviate significantly from those at higher $z$.

### 4.3. Clustering

Next we consider the possibility that Ly$\alpha$ clouds cluster. We adopt a model for uniformly-sized clouds at $z \approx 2$, with clustering. The cloud radius parameter only impacts the hit/miss statistics if it exceeds $S/2$ for one of the sightline pairs considered. We consider two cases, 1) $2R < 39\ h^{-1}$ kpc, and 2) $40\ h^{-1}$ kpc $< 2R < 226\ h^{-1}$ kpc, spanning either none of the sightline gaps or just the smallest. In each case we obtain a best fit for $r_0$ and $\gamma$ of the form $\xi = (r/r_0)^{-\gamma}$ for the two-point correlation function and using all pairs with $R > S/2$.

For case (1), $2R < 39\ h^{-1}$ kpc, a best fit, with $\chi^2/\nu = 0.6$, is obtained for $\gamma = 1.2 \pm 0.3$, $r_0 = (0.61 \pm 0.23)\ h^{-1}$ Mpc, a two-point correlation function that is relatively flat and weak compared to the present day galaxy-galaxy clustering function. For case (2), however, the current data provide little information about the clustering function except that it is very flat, with negative $\gamma$ values and values as large as $\gamma = 1.7$ allowed within the 95% confidence interval.

For a specific model of clustering, the Davis & Peebles (1977) picture of self-similar clustering under gravitational growth, our data correspond to a specific value of the correlation length (here expressed in proper coordinates): at $z \approx 2$, $r_0 = (155 \pm 5)\ h^{-1}$ kpc for case (1), comparable to the value $430(1+z)^{-1}h^{-1}$ kpc found for Ly $\alpha$ clouds along single sightlines (Babul 1991), who assumed small cloud sizes, or $r_0 = (320 \pm 40)\ h^{-1}$ kpc for case (2). The close correspondence between the single and multiple sightline clustering strengths might be seen as evidence for gravitational clustering rather than a cloud size distribution causing the $R(S)$ dependence, since presumably single clouds in single sightlines will produce only one Ly $\alpha$ line (unless there is strong internal velocity substructure). The gravitational clustering hypothesis might also better explain the "multiple hit" phenomenon found in Q1623+268 as described above.

One possible means for distinguishing large, uniform clouds from clusters of smaller ones would be a comparison of the difference in equivalent widths between sightlines (as in Fig. 2), but we conclude on the basis of the current sample that this is inconclusive. From Fig. 4 it is evident that the Q1623+268 Ly $\alpha$ pairs will suffer some contamination due to accidental pairs, even for $\Delta v < 150$ km s$^{-1}$, so we exclude them. In Fig. 6, however, the relative difference of the equivalent widths, $\delta_w \equiv |W_A - W_B|/\max(W_A, W_B)$, for Q0107-025 and Q0307-195 can be compared directly to Fig. 2b. (In Fig. 6, we have included all lines, even those with $W_0 < 0.4$Å, in keeping with Fig. 2b.) While many lines have $W_A \approx W_B$, the deviation of the sample from the $W_A = W_B$ in terms of the average $\delta_w$ is $0.35 \pm 0.043$, compared to $0.28 \pm 0.093$ for uncontaminated lines in Q1343+266. Formally, the result for the Q0107-025/Q0307-195 is more non-uniform than for Q1343+266 and inconsistent with $W_A = W_B$. We express caution, however, since $\delta_w$ for Q0107-025/Q0307-195 is dominated by one pair (from Q0107-025). If it is excluded, giving $\delta_w = 0.30 \pm 0.049$, the results for the two samples are not significantly different. We suggest that a sample several times the size of that presented here might be useful in terms of $W_A - W_B$. In such a case, $\delta_w > (N/2)^{-1/2}$, where $N$ is the number of clouds intercepted in the cluster corresponding to each Ly $\alpha$ "cloud" detection, in each line of sight. If the value of $\delta_w$ is nearly zero in a larger sample, this implies single, large, uniform clouds, or $N$ very large. This is a challenge for future work.

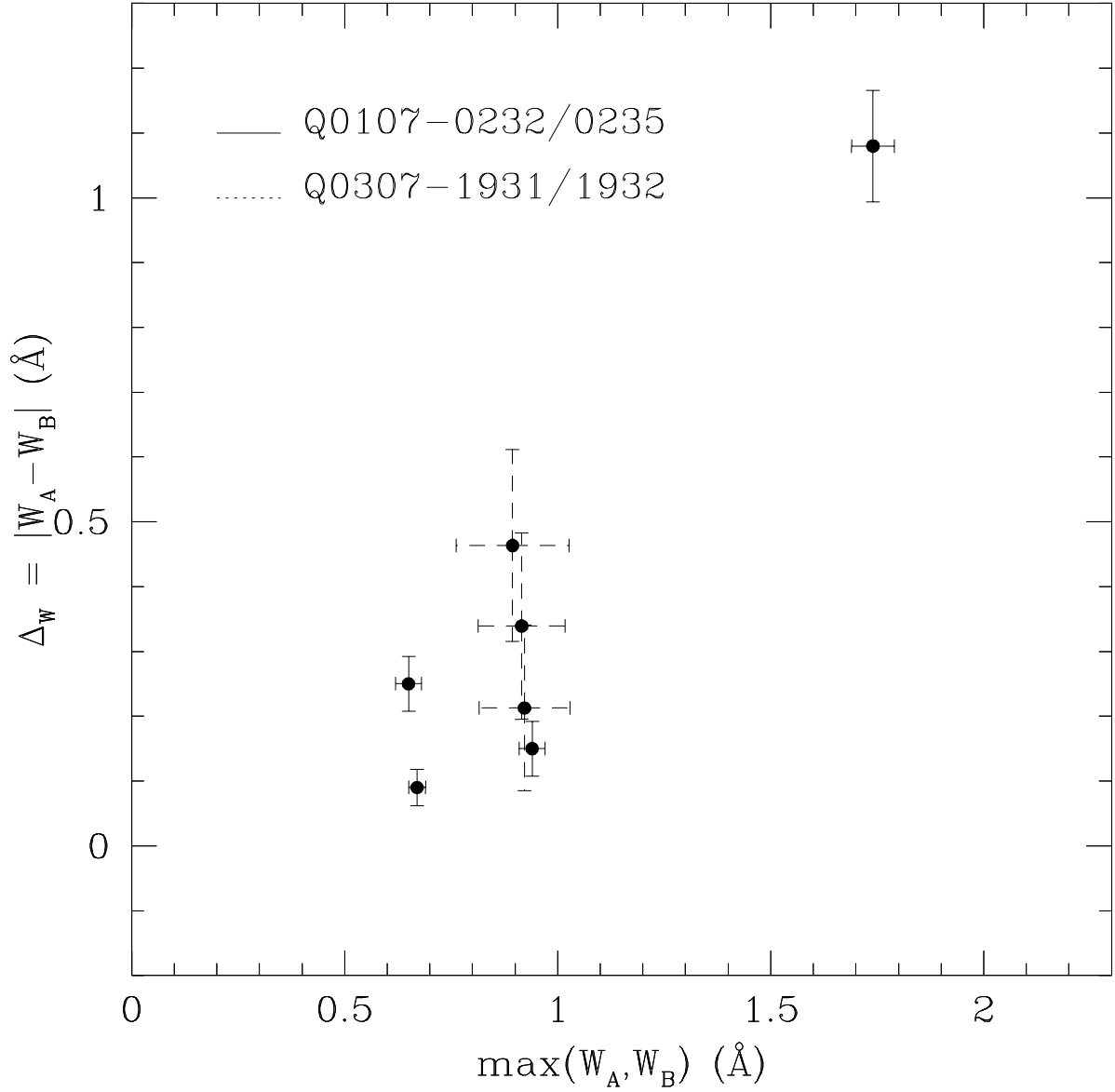

Fig. 6.— The absolute value of the difference of equivalent widths between common lines in the two spectra, $|W_A - W_B|$, is plotted versus the maximum value of $W_A$ or $W_B$, for the QSO pairs Q0107-0232/0107-0235 and Q0307-1931/0307-1932 (the two wider separation pairs with the least evidence of contamination by accidental hits). All detected hits are included, not just those with $W_0 > 0.4$Å.

### 4.4. Filaments

A third possibility, mentioned in §2.1, is that the clouds are elongated or "filamentary". By assuming a randomly oriented uniform cylinders as the model of filamentary structure, this case actually can fit the data fairly well, with $\chi^2/\nu \approx 1.5$ best fit for filaments of 101 $h^{-1}$ kpc radius and any length greater than several Mpc. The 95% confidence interval extends over radii 45 $h^{-1}$ kpc $< r <$ 160 $h^{-1}$ kpc and for any filament length greater than about 1 $h^{-1}$ Mpc. Such a cloud shape is sufficient to produce a large number of hits for close pairs such as Q1343+2640, while insuring a small but significant number for wider pairs. These filamentary clouds would be significantly larger than those predicted in model calculations (Cen et al. 1995). (Note that we are comparing a model at $z = 3$ to data at $z \approx 2$.)

An additional test of the general concept of filamentary clouds is provided by the property that such clouds are usually thought to still be expanding in their longest dimension with a velocity close to that of the Hubble flow. At a redshift of two, the value of the Hubble constant should fall in the range 100 km s$^{-1}$ Mpc$^{-1}$ < $H_0$ < 520 km s$^{-1}$ Mpc$^{-1}$ for the three cosmological cases considered here. For the $q_0 = 1/2$ case (with $H_0 = 520$ km s$^{-1}$ Mpc$^{-1}$), the 150 km s$^{-1}$ velocity scale seen is far below the Hubble scale for the wider pairs. For a low density or $\Lambda$-dominated universe, it is not. Unfortunately, this test is cosmologically model-dependent in an indeterminate way.

Fortunately, pairs at slightly wider separations are easier to find than those studied here, and these may further constrain the filamentary model parameters. Also, triplets of QSO sightlines are very useful for testing the filament hypothesis, since only rarely should an elongated cloud be intersected by all three sightlines, unlike the clustering or distributed cloud-size alternatives. Elsewhere we will reconsider the case of Q1623+268 (Fang & Crotts 1995) and a slightly wider QSO triplet (Crotts et al. 1995) in order to test the filament model in this way.

### 4.5. Concluding Remarks

The data at present are insufficient to distinguish between the clustering and cloud size distribution models, but indicate that the purely unclustered, single-size cloud model developed in §2 cannot explain the behavior of Ly$\alpha$ lines in wider separation pairs. There is also a weak indication that the cloud size distribution or the cloud clustering properties are evolving with redshift. These results call for two further kinds of investigations: further close pair observations at

– 26 –

redshifts other than $z \approx 2$, requiring either close, high-redshift pairs or lower-redshift, bright QSO pairs for $HST$ (both difficult prospects); and more data from pairs with sightline separations in the range of probable cloud radii, 100-600 $h^{-1}$ kpc. The second kind of data will indicate whether the transition from high hit fraction, $\psi \approx 1$ to $\psi \approx 0$ is abrupt and therefore consistent primarily with weakly clustered clouds of a single size, or gradual, as might be indicated by a smooth distribution in cloud radii. Several such pairs are known to exist and are candidates for further study. As mentioned previously, triplets or group of QSOs with proper separations $\approx$ 1 Mpc are also useful in determining whether the clouds are more like filaments versus spheres or disks.

We thank Mark Fardal for pointing out the difference between $\phi$ and $\psi$. We thank the referee, Ray Weymann, for many helpful comments. This work was supported by NSF grant AST 90-58510, NASA grant number AR-05785.01-94A, and a gift from Sun microsystems (to JB), AST 91-16390, the David and Lucile Packard Foundation and NSF grant AST 90-22586 (to AC), and NSF AST 90-20757 (for RD).